\documentclass[trackchanges,twocolumn]{aastex701}

\usepackage{amsmath}
\usepackage{amssymb}	
\usepackage{natbib}
\usepackage{hyperref}
\hypersetup{colorlinks,linkcolor={blue},citecolor={blue},urlcolor={blue}}  
\usepackage[figuresright]{rotating}
\usepackage{threeparttable}
\usepackage{subfigure}
\usepackage{soul}
\usepackage[normalem]{ulem}
\usepackage{cancel}
\usepackage{graphicx}
\graphicspath{ {./figure} }

\newcommand{\cmcu}{{cm$^{-3}$}}
\newcommand{\s}{{s$^{-1}$}}
\newcommand{\kms}{km\,s$^{-1}$}

\newcommand{\ciii}{\ion{C}{3}]}

\newcommand{\civ}{\ion{C}{4}}
\newcommand{\lya}{Ly$\alpha$}
\newcommand{\heii}{\ion{He}{2}}
\newcommand{\oii}{[\ion{O}{2}]}
\newcommand{\oiii}{\ion{O}{3}]}
\newcommand{\foiii}{[\ion{O}{3}]}
\newcommand{\oiiiuv}{\ion{O}{3}]}

\newcommand{\ha}{H$\alpha$}
\newcommand{\hb}{H$\beta$}
\newcommand{\hg}{H$\gamma$}
\newcommand{\hd}{H$\delta$}

\newcommand{\nii}{[\ion{N}{2}]}

\newcommand{\niv}{\ion{N}{4}]}

\newcommand{\neiii}{[\ion{Ne}{3}]}
\newcommand{\nev}{[\ion{Ne}{5}]}

\newcommand{\lae}{C3PO-84478}
\newcommand{\jwst}{\textit{JWST}}
\newcommand{\hst}{\textit{HST}}

\begin{document}

\title{C3PO: A \lya\ Emitting Galaxy at $z\sim9$}

\author[0000-0003-3424-3230]{Weida Hu}
\email{weidahu.astro@gmail.com}
\affiliation{Department of Physics and Astronomy, Texas A\&M University, College Station, TX 77843-4242, USA}
\affiliation{George P. and Cynthia Woods Mitchell Institute for Fundamental Physics and Astronomy, Texas A\&M University, College Station, TX 77843-4242, USA}

\author[0000-0001-7503-8482]{Casey Papovich}
\email{papovich@tamu.edu}
\affiliation{Department of Physics and Astronomy, Texas A\&M University, College Station, TX 77843-4242, USA}
\affiliation{George P. and Cynthia Woods Mitchell Institute for Fundamental Physics and Astronomy, Texas A\&M University, College Station, TX 77843-4242, USA}

\author[orcid=0000-0001-6251-4988,sname='Hutchison']{Taylor A. Hutchison} 
\altaffiliation{NASA Postdoctoral Fellow}
\affiliation{Astrophysics Science Division, NASA Goddard Space Flight Center, 8800 Greenbelt Rd, Greenbelt, MD 20771, USA}
\email{taylor.hutchison@nasa.gov}

\author[orcid=0000-0003-2366-8858,sname='Larson']{Rebecca L.\ Larson}
\altaffiliation{Giacconi Postdoctoral Fellow}
\affil{Space Telescope Science Institute, 3700 San Martin Drive, Baltimore, MD 21218, USA}
\email{rlarson@stsci.edu}

\author[0000-0002-7959-8783]{Pablo Arrabal Haro}
\email{parrabalh@gmail.com}
\affiliation{Center for Space Sciences and Technology, UMBC, 5523 Research Park Dr, Baltimore, MD 21228 USA }
\affiliation{Astrophysics Science Division, NASA Goddard Space Flight Center, 8800 Greenbelt Rd, Greenbelt, MD 20771, USA}

\author[0000-0001-8519-1130]{Steven L. Finkelstein}
\affiliation{Department of Astronomy, The University of Texas at Austin, Austin, TX, USA}
\affiliation{Cosmic Frontier Center, The University of Texas at Austin, Austin, TX, USA}
\email{stevenf@astro.as.utexas.edu}

\author[0000-0001-7151-009X]{Nikko J. Cleri}
\affiliation{Department of Astronomy and Astrophysics, The Pennsylvania State University, University Park, PA 16802, USA}
\affiliation{Institute for Computational and Data Sciences, The Pennsylvania State University, University Park, PA 16802, USA}
\affiliation{Institute for Gravitation and the Cosmos, The Pennsylvania State University, University Park, PA 16802, USA}
\email{cleri@psu.edu}

\author[0000-0001-9495-7759]{Lu Shen}
\email{lushen@tamu.edu}
\affiliation{Department of Physics and Astronomy, Texas A\&M University, College Station, TX 77843-4242, USA}
\affiliation{George P. and Cynthia Woods Mitchell Institute for Fundamental Physics and Astronomy, Texas A\&M University, College Station, TX 77843-4242, USA}

\author[0000-0001-7593-9205]{Jiayang Yang}
\email{annabellayang@tamu.edu}
\affiliation{Department of Physics and Astronomy, Texas A\&M University, College Station, TX 77843-4242, USA}
\affiliation{George P. and Cynthia Woods Mitchell Institute for Fundamental Physics and Astronomy, Texas A\&M University, College Station, TX 77843-4242, USA}

\begin{abstract}
We present deep, $>10$~hr, \jwst\ NIRSpec G140M and G395M observations of a previously known galaxy \lae\ at $z=8.99781$.
Our new NIRSpec data from the Carbon-3 plus Oxygen (C3PO) survey detect \lya\ emission with an escape fraction of $1.5\pm0.3\%$ and a velocity offset of $210\pm80$ \kms, making \lae\ the third \lya\ emitting galaxy confirmed at $z\gtrsim 9$.
The galaxy has a low stellar mass of $2.2\pm0.1\times 10^7\ M_\odot$, a young stellar age of $1.6\pm0.2$ Myr, and a low oxygen abundance of $12+\log(\mathrm{O/H}) = 7.50\pm0.08$. 
UV diagnostic diagrams incorporating the \civ, \oiiiuv, and \ciii\ emission lines indicate that the ionization of \lae\ is dominated by the recent starburst.
The low \lya\ escape fraction of \lae\ implies that \lae\ resides in a small ionized bubble with radius $R \sim 0.07$ to 0.20 pMpc, which can be produced by \lae\ alone. 
We find no evidence for either a large-scale galaxy overdensity around \lae\ or a significant contribution from nearby galaxies to its ionizing photon budget. 
Taken together, these results suggest that \lae\ likely represents an ionized bubble powered by a single low-mass, vigorously star-forming galaxy during the early stages of cosmic reionization.
\end{abstract}


\section{Introduction} 

Cosmic reionization marks a critical phase transition of the intergalactic medium (IGM) in the history of the Universe. 
During this epoch, early star-forming galaxies and active galactic nuclei (AGN) emitted copious ionizing photons into their surrounding IGM, progressively ionizing the hydrogen gas and transforming the IGM from a predominantly neutral state to a mostly ionized one \citep[e.g.,][]{Barkana2001,Fan2006a,Finkelstein2019,Madau2024}. 
Over the past three decades, substantial progress has been made in constraining the timeline of cosmic reionization.
Observations of the Gunn--Peterson trough \citep{Gunn1965} in quasar spectra indicate that reionization has been largely complete by $z \sim 5$ -- 6 \citep[e.g.,][]{Fan2006b,Becker2015,Eilers2018,Yang2020,Zhu2021}. 
Measurements of the Thomson optical depth from cosmic microwave background (CMB) polarization suggest that the midpoint of reionization occurs at $z = 7.7 \pm 0.8$ \citep{PlanckCollaboration2020}, and the reionization begins before this epoch.
Furthermore, statistical studies based on the \lya\ properties of galaxies indicate a rapid increase in the neutral hydrogen fraction of the IGM from $z \sim 5$ to $z \gtrsim8$ \citep{Malhotra2004,Tilvi2014,Ouchi2010,Zheng2017,Mason2018,Hu2019,Morales2021,Wold2022, Bolan2022,Ning2022,Tang2024,Napolitano2024}.

As \lya\ is a resonant transition of hydrogen, even a small amount of residual neutral hydrogen in the IGM can effectively scatter \lya\ photons out of sightline, leading to significant attenuation in the observed \lya\ luminosity of the galaxy.
This enables \lya\ a particular sensitive probe of reionization.
The detection of \lya\ emission implies the presence of a locally ionized bubble surrounding the galaxy, within which \lya\ photons can redshift out of resonance before encountering the neutral IGM, thereby reducing IGM attenuation \citep{Furlanetto2004,Dijkstra2014,Mason2020}.
This enables LAEs as powerful tracers of ionized bubbles \citep{Hayes2023b,Lu2025,Nikolic2025}.
Observations of \lya\ emitting galaxies (LAEs) have revealed numerous LAE overdensities at $z \sim 6$ -- 8 \citep[e.g.,][]{Jiang2018,Tilvi2020,Hu2021,Endsley2022,Jung2022,Witstok2024,Hu2026}, indicating that the ionized bubbles produced by individual galaxies have already overlapped at these redshifts, collectively forming large ionized bubbles surrounding overdense structures. 

However, the emerging picture of ionized bubbles at higher redshifts ($z>8$), when the universe is largely neutral, remains elusive. 
Prior to \jwst, only two LAEs were known at $z > 8$, separated by $\sim 3.5$ physical Mpc \citep[pMpc;][]{Zitrin2015,Larson2022}. 
Several independent studies using \hst\ and \jwst\ imaging identified a galaxy overdensity surrounding these
sources \citep{Larson2022,Leonova2022,Whitler2024,Whitler2025}, suggesting that large, overlapping ionized bubbles may already exist at $z>8$.
Nonetheless, \citet{Witstok2024} show evidence for the diversity of ionized bubble properties at these early epochs based on the \jwst\ NIRSpec observation of three LAEs at $z>8$. 
Although a bright LAE at $z \sim 8.28$ in \citet{Witstok2024} likely occupies a substantially larger ionized bubble that requires additional contributions from nearby companion galaxies, the two other LAEs may reside within very small ionized bubbles produced primarily by themselves.
These observations suggest that at $z>8$ we could be witnessing the earliest stages of bubble formation, offering a unique opportunity to investigate which types of galaxies are responsible for generating the very first ionized regions in the Universe.

At lower redshifts, LAEs exhibit diverse ionizing sources, including star formation and active galactic nuclei (AGNs) \citep{Sobral2018}.
However, at $z>8.5$, an increasing fraction of LAEs appear to host AGNs \citep{Larson2023,Bunker2024,Maiolino2024,Morishita2025,Witstok2024}.
For example, one of the most distant known LAEs, GN-z11, at $z=10.6$, exhibits very high-ionization emission lines and a broad blueshifted \civ\ absorption trough, strongly suggesting the presence of an AGN \citep{Bunker2023,Maiolino2024}.
A little red dot (LRD) at $z=8.63$ has also been reported to show a strong \lya\ emission line \citep{Morishita2025}. 
Although \citet{Witstok2024} reported an LAE at $z=8.72$ to be likely dominated by a very recent, vigorous star-formation burst, it also shows evidence of an AGN.
Together, these findings raise the question of whether AGNs play a dominant role in initiating the formation of ionized bubbles at the highest redshifts.
Addressing this question requires discovering more LAEs at the earliest stages of reionization, as well as detailed constraints on their intrinsic ionizing properties and their surrounding environments.

In this work, we report the detection of a \lya\ emission line in a $z=9$ galaxy, \lae\ (previously known as CEERS-24). 
\lae\ was photometrically selected as a $z\gtrsim8.5$ galaxy candidate from the CEERS NIRCam imaging \citep{Finkelstein2023}.
It was then observed with NIRSpec medium-grating spectroscopy with configurations G140M/F100LP, F235M/F170LP, and G395M/F290LP as part of the CEERS program.
The detection of \foiii~$\lambda\lambda4959,5007$ by the CEERS program confirms its redshift to be $z=8.998$ \citep{Fujimoto2023,Tang2024}; however, no other line was significantly detected.
We carried out a NIRSpec spectroscopy program, Carbon III Plus Oxygen survey (C3PO; GO-5943; PIs: Papovich, Hu, Hutchison), which uses deep  G140M/F100LP and G395M/F290LP to obtain deep rest-frame UV and optical spectra of galaxies at $5.6<z<9$, including \lae.
Our NIRSpec G140M spectrum detects the \lya\ emission line of \lae, making it the third object at $z\gtrsim9$ with confirmed \lya\ emission.  Here we use the deep NIRSpec data to characterize the physical properties of this galaxy and to estimate its impact on its surroundings. 

This paper is structured as follows. Section \ref{sec:obs} describes the NIRCam imaging and NIRSpec spectroscopy data used in this work, and 
Section \ref{sec:meas} presents our measurements and analysis of \lae. 
In Section \ref{sec:disc}, we discuss the ionizing source, ionized bubble, and environment of \lae.
Finally, we summarize our results in Section \ref{sec:summ}.
Throughout this study, we adopt the Planck 2018 cosmological parameters \citep{PlanckCollaboration2020}: $\Omega_m=0.3111$, $\Omega_\Lambda=0.6889$ and $H_0=67.66$ \kms\ Mpc$^{-1}$, where $\Omega_m$ and $\Omega_\Lambda$ are the densities of total matter and dark energy and $H_0$ is the Hubble constant.

\section{Observations} \label{sec:obs}
In this section, we present the NIRCam imaging observation and NIRSpec MSA medium-grating spectroscopy observation utilized to study \lae. 

\subsection{NIRCam Imaging}
The NIRCam imaging observations utilized in this work were taken as part of the Cosmic Evolution Early Release Science (CEERS) Survey \citep{Finkelstein2025}. 
The CEERS is an \jwst\ early release science program to observe the Extended Groth Strip (EGS) field using the four modes of \jwst\ operations: NIRCam imaging, NIRCam grism spectroscopy, NIRSpec  MSA spectroscopy, and MIRI imaging. 
The NIRCam imaging observations include 10 pointings using F115W, F150W, F200W, F277W, F356W, F410M, and F444W filters.
The data was reduced using the standard \jwst\ pipeline \citep{Bushouse2025} with some customized steps to remove snowballs, wisps, and $1/f$ noise. 
A full description of the CEERS survey and data reduction can be found in \citet{Bagley2023} and \citet{Finkelstein2025}.

We use photometric catalogs from the UNICORN project (S. Finkelstein et al. in prep). These catalogs are broadly based on the photometric procedures of \citep{Finkelstein2024}, thus we refer the reader there for more information. In brief, these catalogs are Source Extractor based \citep{Bertin1996}, using F277W+F356W as the detection image, and provide robust estimates of object colors, total fluxes, and photometric uncertainties. Photometric redshifts are included, using Lazy \footnote{https://github.com/hollisakins/Lazy.jl}, which is a Julia-based version of the template-based fitting code EaZY \citep{Brammer2008}.


\subsection{NIRSpec MSA Spectroscopy}
The NIRSpec spectroscopy utilized in this work is taken through the C3PO program, which primarily targets galaxies at $5.6<z<9$ in the CEERS field to obtain high signal-to-noise ratio (S/N) rest-frame UV and optical spectra.
The C3PO NIRSpec observation consists of two MSA pointings, each observed with the G140M/F100LP and G395M/F290LP grating/filter pairs.
For each pointing, we design four MSA configurations where the high-priority objects remain on each confirmation, but with lower-priority objects being replaced. 
Each MSA configuration is carried out using a three-shutter slitlet with nodding between the three shutters.
We also manually open shutters that are adjacent to the stuck-open shutters to maximize observing efficiency. 
The effective exposure time of each MSA configuration is $\sim11.9$ ks and $\sim9.3$ ks for G140M/F100LP and G395M/F290LP, respectively. 
Targets may be observed on one to four MSA configurations, depending on their priority, such that the total effective exposure time for high-priority targets can reach $\sim47.8$ ks for G140M/F100LP and $\sim37.1$ ks for G395M/F290LP.

The NIRSpec data are reduced using the standard \jwst\ pipeline \citep[version 1.18;][]{Bushouse2025} with the Calibration Reference Data System (CRDS) version jwst\_1350.pmap.
We include several modifications to mitigate the unrealistically large uncertainties in the \texttt{fflat} reference file and to extend wavelength coverage of G140M/F100LP configuration to 1.93 $\mu$m. 
To extend the wavelength coverage of G140M/F100LP, we modify reference files including \texttt{wavelengthrange}, \texttt{fflat}, \texttt{sflat}, \texttt{photom}, and \texttt{apcorr}.
In the \texttt{wavelengthrange} reference file, we set the valid wavelength interval for G140M/F100LP to [0.97, 1.93] $\mu$m. 
We extrapolate the fore-optics flat field (\texttt{fflat}) and the spectrograph flat field (\texttt{sflat}) from 1.89 $\mu$m to 1.93 $\mu$m using a linear extrapolation.
Because the instrumental responses in the \texttt{photom} and \texttt{apcorr} reference files are ones, we extend the wavelength grid in those reference files to 1.93 $\mu$m with ones.
Then, we extract the 1D spectra using an optimal extraction algorithm introduced in \citet{Horne1986}.

Details of the target selection and observation strategy will be described in Papovich et al. in prep. 

\begin{figure*}
    \centering
    \includegraphics[width=\linewidth]{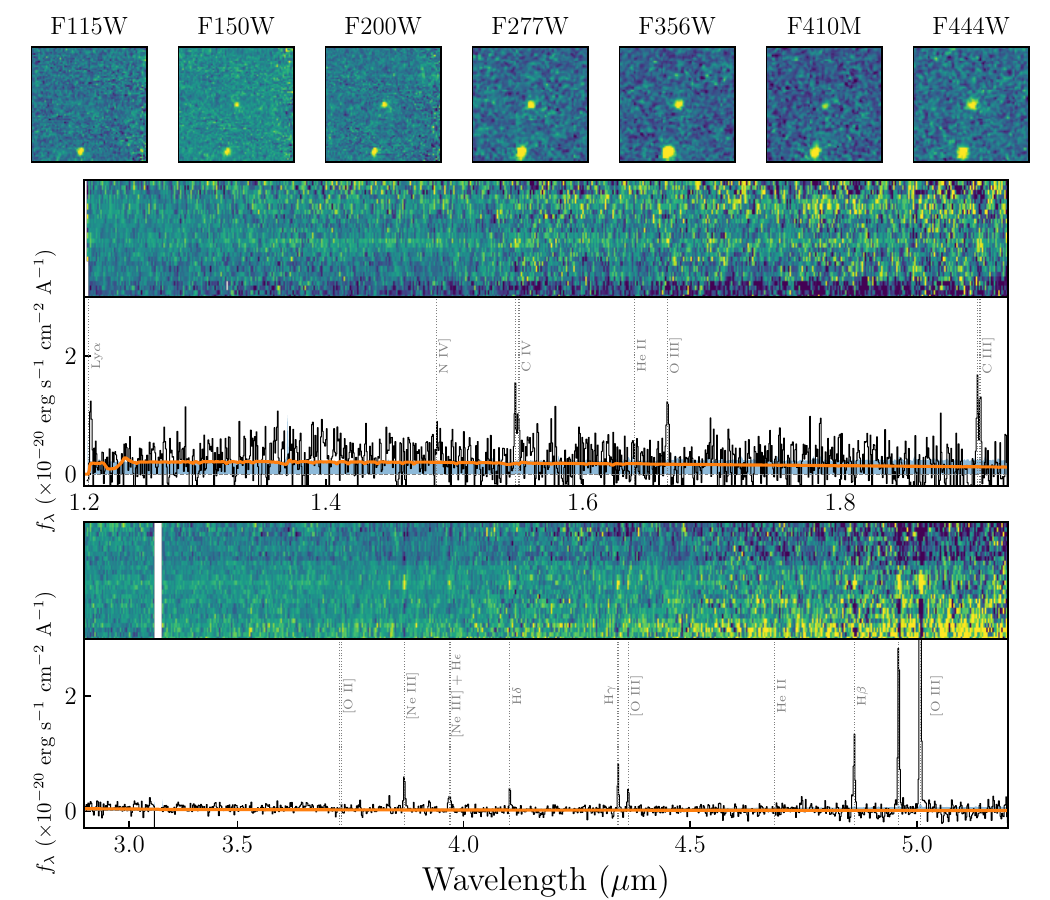}
    \caption{\jwst\ NIRCam imaging and NIRSpec spectroscopy of \lae. The top panel shows the NIRCam images, the middle panel shows the NIRSpec G140M spectrum, and the bottom panel shows the NIRSpec G395M spectrum. In the bottom two panels, the black curves represent the 1D spectra and the blue shades represent the error spectra. The gray vertical dotted lines highlight emission lines frequently detected in high-redshift galaxies. We also present our best-fit continuum model from SED modeling as the solid orange curves. The best-fit SED model is presented in Figure \ref{fig:bestsed}.}
    \label{fig:fullspec}
\end{figure*}

\section{Measurements and Results} \label{sec:meas}

\begin{deluxetable}{l c c}
    \tablecaption{Emission Line Fluxes in \lae  \label{tab:flux}}
    \tablehead{\colhead{Line} & \colhead{$f_\mathrm{line}$} & \colhead{EW$_0$} \\ 
    \colhead{} & \colhead{$\times10^{-20}$ erg s$^{-1}$ cm$^{-2}$} & \colhead{\AA}}
    \startdata
    \lya & $19.4\pm3.1$ & $6.8\pm1.1$ \\
    \civ\ $\lambda\lambda1548,1550$ & $37.3\pm5.2$ & $19.8\pm2.7$ \\
    \heii & $<6.4\ (2\sigma)$ & $<3.9$ \\
    \oiii~$\lambda1666$ & $24.8\pm4.0$ & $15.2\pm2.5$ \\
    \ciii~$\lambda1907$ & $23.0\pm3.1$ & $20.7\pm2.8$ \\
    \ciii~$\lambda1909$ & $19.8\pm3.1$ & $17.8\pm2.8$ \\
    \ciii~$\lambda\lambda1907,1909$ & $42.8\pm4.3$ & $38.6\pm3.9$\\
    \oii~$\lambda\lambda3727,3729$ & $<3.6\ (2\sigma)$ & $<14.2$ \\
    \neiii\ $\lambda3869$ & $27.2\pm1.8$ & $108.4\pm7.2$\\
    \hd & $16.4\pm1.8$ & $77.7 \pm 8.5$\\
    \hg & $28.0\pm1.9$ & $150.1\pm 10.2$\\ 
    \foiii\ $\lambda4363$ & $16.3\pm1.8$ & $87.4\pm9.7$ \\
    \hb & $53.7\pm3.2$ & $370.3\pm22.1$ \\
    \foiii\ $\lambda4959$ & $126.6\pm3.3$ & $904.3\pm23.6$ \\
    \foiii\ $\lambda5007$ & $375.2\pm3.8$ & $2680.0\pm27.1$ \\
    \enddata
\end{deluxetable}

\subsection{Redshift Determination and Emission Line Measurements}

\begin{figure*}
    \centering
    \includegraphics[width=0.9\linewidth]{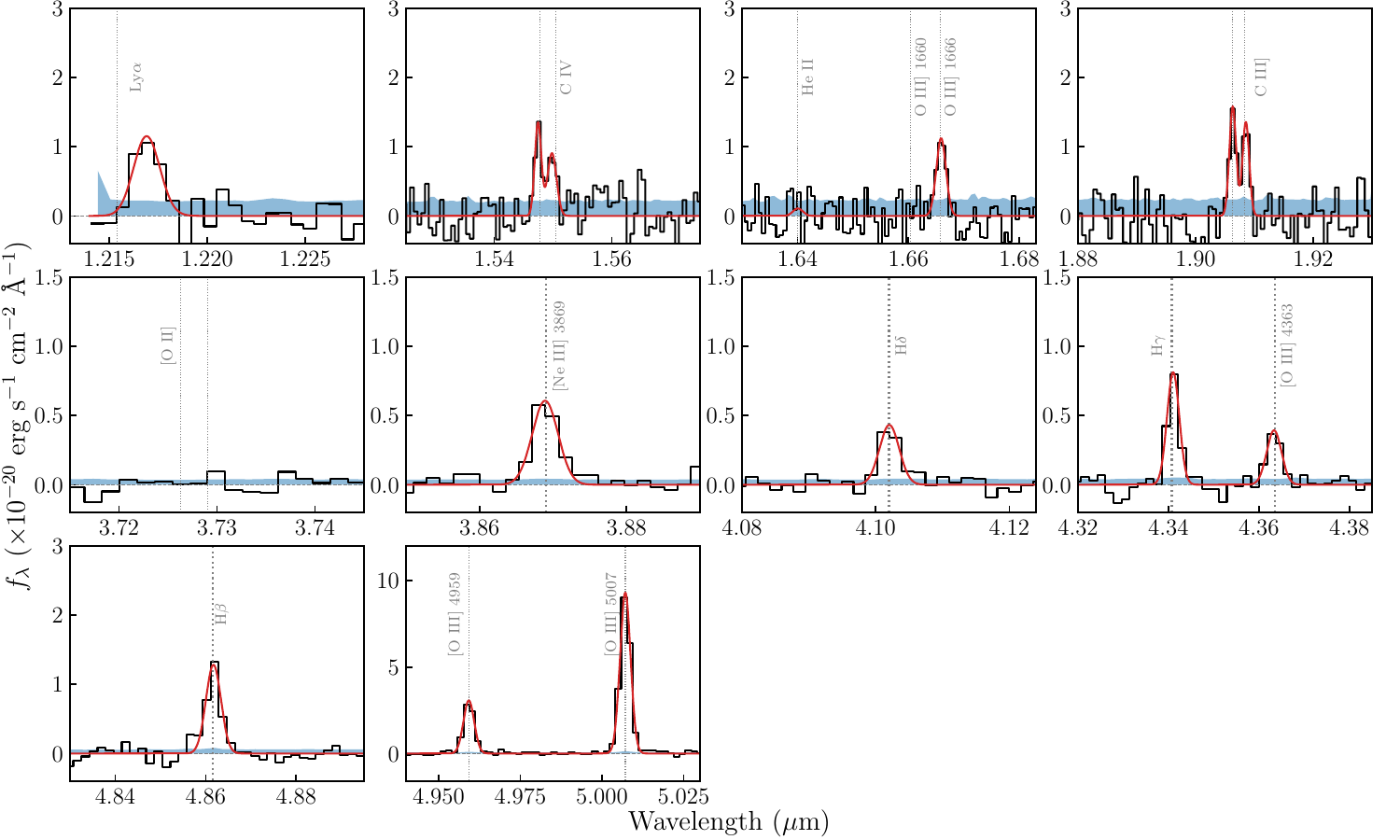}
    \caption{Continuum-subtracted 1D spectra of the emission lines seen in \lae. In each subfigure, we present the 1D spectrum, the $1\sigma$ error spectrum, and the best-fit model, as the black curves, blue shades, and red curves, respectively.}
    \label{fig:emline}
\end{figure*}

Figure \ref{fig:fullspec} shows the 2D and 1D spectra of G140M and G395M of \lae. 
The G140M/F100LP spectrum reveals marginal detection of UV continuum, while both G140M and G395M show significant detection of multiple emission lines.
We highlight the positions of typically strong rest-frame UV and optical emission lines based on their vacuum wavelengths in Figure \ref{fig:fullspec}, including \lya, \niv, \civ~$\lambda\lambda1548,1550$, \oiii~$\lambda1666$, \ciii~$\lambda\lambda1907,1909$, \oii~$\lambda\lambda3727,3729$ \neiii~$\lambda3869$, \foiii~$\lambda4363$, \foiii~$\lambda\lambda4959,5007$, and several hydrogen Balmer lines (\hg, \hd, \hb, etc.). 

To determine the spectroscopic redshift of \lae, we use \texttt{emcee} \citep{Foreman-Mackey2013} to simultaneously fit \foiii~$\lambda\lambda$4959,5007 and \hb\ emission lines, assuming they emerge from the same region (i.e., have the same redshift).
We adopt a single Gaussian profile with a constant underlying continuum for each line.
We obtain the redshift to be $8.99781\pm0.00005$, consistent with those derived in \citet{Tang2023} and \citet{Fujimoto2023} using the CEERS NIRSpec G395M spectrum.
Given the redshift uncertainty ($\Delta v=1.5$ \kms) is considerably small compared to the spectral resolution ($R\sim1000$), we adopt this value as the systemic redshift for \lae\ and do not consider redshift uncertainty in the following analysis. 

We measure the flux of the emission line by fitting a Gaussian profile to the continuum-subtracted spectrum.
To derive the continuum-subtracted spectrum, we utilize the best-fit SED model from \texttt{BAGPIPES} (see Section \ref{sec:sed}) with the emission lines removed.
This method enables us to account for the contribution of stellar absorption features, such as \civ\ and hydrogen Balmer.
For single lines, including \foiii~$\lambda4959$, \foiii~$\lambda5007$, \neiii~$\lambda3869$, \oiii~$\lambda1666$, \heii~$\lambda1640$, \lya~$\lambda1216$, and hydrogen Balmer lines, we adopt a single-Gaussian model. 
For blended emission lines, including \civ~$\lambda\lambda1548,1550$, \ciii~$\lambda\lambda1907,1909$, and \oii~$\lambda\lambda3727,3729$, we adopt a double-Gaussian profile. 
As the \oii~$\lambda\lambda$3727,3729 and \heii~$\lambda1640$ lines are not detected in the spectrum, we fix the line widths of the \oii\ doublets to match those of \neiii~$\lambda3869$, and set the \heii\ line width to match that of \oiii~$\lambda$1666.
In addition, we adopt a skewed Gaussian profile to fit the \lya\ line, because it can be resonantly scattered by the ISM \citep[e.g.,][]{Verhamme2006}, resulting in a redshifted, asymmetric line profile.

We use a Monte Carlo approach to calculate the flux uncertainties. 
Specifically, we perturb the line profiles based on their error spectra. We generate 1000 realizations for each line and refit them using the same fitting procedure. 
We take the standard deviation of the flux distribution as the measurement uncertainty.
We present the best-fit models in Figure \ref{fig:emline} and report the measurements in Table \ref{tab:flux}.
For \oii~$\lambda\lambda$3727,3729 and \heii~$\lambda1666$, we report their $2\sigma$ upper limits.

\begin{deluxetable}{l r}
    \tablecaption{Galaxy Properties of \lae  \label{tab:galprop}}
    \tablehead{\colhead{Property} & \colhead{Value}}
    \startdata
    $z_\mathrm{spec}$ & $8.99781\pm0.00005$ \\
    R.A. & 14:19:35.336  \\
    Dec. & 52:50:37.89 \\
    $r_\mathrm{eff}$ & $100\pm63$ pc\\
    \hline
    \multicolumn{2}{c}{SED}\\
    \hline
    M$_\mathrm{UV}$ & $-19.33\pm0.04$ \\
    $\beta_\mathrm{UV}$ & $-2.04\pm0.11$ \\
    $\log\ (M_\star/M_\odot)$ & $7.35\pm0.02$\\  
    $f_{\rm LyC}^{\rm esc}$ & $25^{+3}_{-4}\%$ \\
    $E(B-V)_\mathrm{SED}$ & $0.00^{+0.00}_{-0.00}$ \\
    age & $1.6\pm0.2$ Myr \\
    SFR & $14.2^{+2.5}_{-2.3}\ M_\odot\ \rm yr^{-1}$ \\
    $\Sigma_\mathrm{SFR}$ & $\geq142^{+25}_{-23}\ M_\odot$ yr$^{-1}$ kpc$^{-2}$\\
    $\log U_\mathrm{SED}$ & $-0.03^{+0.02}_{-0.03}$\\
    \hline
    \multicolumn{2}{c}{Spectrum}\\
    \hline
    $E(B-V)_\mathrm{nebular}$$^a$ & $<0.03\ (1\sigma)$ \\
    $n_e(\mathrm{C^{2+}})$ & $12900^{+12500}_{-8600}$ cm$^{-3}$ \\
    $T_e(\mathrm{O^{2+}})^b$ & $23600^{+1800}_{-1700}$ K \\
    $\log\, U$ & $\geq -1.1$ \\
    $12+\log(\mathrm{O/H})$ & $7.50\pm0.08$ \\
    $\log\ \mathrm{(C/O)}$ & $\geq -0.71\pm0.09$ \\
    \hline
    \multicolumn{2}{c}{\lya} \\
    \hline
    $\Delta v_\mathrm{Ly\alpha}$$^c$ & $210\pm80$ \kms\\
    $f^\mathrm{esc}_\mathrm{Ly\alpha}\ (=f^\mathrm{int,esc}_\mathrm{Ly\alpha}\times T_\mathrm{IGM})$ & $1.5\pm0.3\%$ \\
    \enddata
    \tablecomments{
    \begin{itemize}
    \item[$^a$] $E(B-V)_\mathrm{nebular}$ is measured using the \hb/\hg\ line ratio, assuming a \citet{Calzetti2000} attenuation law. 
    \item[$^b$] The temperature of O$^{2+}$ zone is derived assuming an electron density $n_e=500$.
    \item[$^c$] The \lya\ velocity offset is measured by fitting \lya\ line with a skewed Gaussian profile. We adopt half a spectral pixel as the velocity uncertainty, corresponding to $\sim80$ \kms.
    \end{itemize}}
\end{deluxetable}

\subsection{Stellar Population} \label{sec:sed}

We infer the stellar population properties of \lae\ by jointly modeling its photometry and C3PO spectroscopy with the Python package \texttt{BAGPIPES} v1.3.6 \citep{Carnall2018}, adopting the BPASS stellar population models \citep{Eldridge2016,Stanway2016} together with the ionizing photon escape model of \citet{Giovinazzo2026}. 
By default, \texttt{BAGPIPES} assumes that all ionizing photons are absorbed by the ISM and reprocessed into nebular emission.
The ionizing photon escape model of \citet{Giovinazzo2026} instead assumes a picket-fence ISM, in which ionizing photons escape through low-column-density, dust-free channels (see a similar work by \citealt{Papovich2025}). 
The escape fraction of ionizing photons is set by the covering fraction of these channels, while the nebular emission is produced by the fraction $(1-f_{\rm esc})$ of ionizing photons absorbed by the high-column-density ISM.

Because light from young massive stars can outshine the older stellar population, integrated SED fitting may underestimate the stellar mass \citep[e.g.,][]{Papovich2001,Narayanan2024}.
To evaluate the contribution of old stellar population, we perform two fittings with different parameterizations of the star formation history (SFH).
First, we utilize the ``continuity'' prior \citep{Leja2019} with six lookback-time
bins of 0 -- 3, 3 -- 10, 10 -- 50, 50 -- 100, 100 -- 200, and 200 -- 500 Myr, similar to
those adopted in previous studies \citep[e.g.,][]{Endsley2024,Begley2025};
the SFR is constant within each bin.
While this parameterization can capture mass formed at early times, the fixed
bin widths may overestimate the duration of the recent starburst.
In particular, the best-fit SFH of \lae\ places all star formation in the most recent (0--3 Myr) bin, with negligible contribution from earlier bins.
Motivated by this result, we utilize a constant SFH as the second
parameterization to better constrain the young stellar population.
The results of this fitting is adopted for the following analysis.

In both fittings, we allow the ionizing photon escape fraction $f^{\rm esc}_{\rm LyC}$ to vary between 0 and 1.
The stellar metallicity is allowed to vary between 0.001 and 1 $Z_\odot$.  
We include nebular emission with the gas-phase metallicity equal to that of the stellar populations, and an ionization parameter, $\log U$, in the range of $-4$ to $0$. 
We also allow the emission lines to be broadened by an intrinsic velocity dispersion ranging from 0 to 300 \kms.

For dust attenuation, we adopt a flexible modification of the Calzetti dust law \citep{Calzetti2000}, as formulated by \citet{Salim2018}.
This parametrization has two free parameters: $\delta$, quantifying the deviation from the slope of the Calzetti law, and $B$, representing the strength of the 2175 \AA\ absorption feature.
When $\delta=0$, the Salim dust law is equivalent to the Calzetti dust law.
We fix $B=0$, as the 2175 \AA\ absorption feature is rarely observed in high-redshift galaxies \citep{Witstok2023}.
We allow $A_V$ and $\delta$ to vary in the range of 0 -- 1 and $-10$ -- 10, respectively.

We further account for instrumental effects of NIRSpec, including its wavelength-dependent spectral resolution and slitloss.
To model the variable spectral resolution of NIRSpec in the SED fitting, we utilize the resolution models for the G140M and G395M spectra from the \textit{JWST} User Documentation\footnote{\url{https://jwst-docs.stsci.edu/jwst-near-infrared-spectrograph/nirspec-instrumentation/nirspec-dispersers-and-filters}}.
To account for the slitlosses and calibration uncertainties of NIRSpec spectroscopy, we adopt a multiplicative scaling factor (assumed to be a second-order Chebyshev polynomial) in \texttt{BAGPIPES} to estimate the wavelength-dependent flux calibration to the broadband photometry.
We allow the zeroth order to vary from 0.1 to 10, while the first and second orders are allowed to vary from $-0.5$ to 0.5. 

We report the properties derived from SED fitting in Table \ref{tab:galprop} and present the best-fit SED in Figure \ref{fig:bestsed}.

\subsection{Galaxy Size and Star Formation Rate Surface Density} 

\lae\ exhibits a compact morphology.
To determine the effective radius of \lae, we use the Python package \texttt{Pysersic} \citep{Pasha2023} to model the two-dimensional profile of the galaxy in the F150W image, which provides the highest spatial resolution compared with other NIRCam images at longer wavelengths.
We fit a S\'ersic profile accounting for the instrumental point spread function (PSF) provided by the CEERS collaboration \citep{Cox2025}.
Because \lae\ is detected at relatively low S/N, the measured ellipticity is highly sensitive to random background fluctuations, which could lead to spurious elongations.
To mitigate this effect, we assume a circular profile of the target (i.e., ellipticity $=0$) during the fitting.
We derive an effective radius $r_\mathrm{eff}=0.022\pm0.014$ arcsec, corresponding to a physical size of $100\pm63$ pc. 
We also perform the same procedure on other NIRCam filters and obtain the effective radius consistent with that of F150W within $1\sigma$ uncertainty. 
The large uncertainty of size measurement suggests \lae\ is nearly unresolved in the NIRcam images.
Therefore, we adopt the $2\sigma$ upper limit $<126$ pc in the following analysis.

We measure the star formation rate surface density $\Sigma_\mathrm{SFR}$ using the formula:
\begin{equation}
    \Sigma_\mathrm{SFR} = \frac{\mathrm{SFR}}{2\pi r^2_\mathrm{eff}},
\end{equation}
and we obtain $\Sigma_\mathrm{SFR} \geq 142^{+25}_{-23}\ M_\odot$ yr$^{-1}$ kpc$^{-2}$.

\subsection{ISM Properties} \label{sec:ism}

With abundant nebular emission lines in \lae, we measure the IGM properties in this section, including nebular dust attenuation, electron density and temperature, ionization parameter, and metal abundances.

\vspace{0.5em}
\textit{Nebular dust attenuation:} 
We measure the nebular dust attenuation using the Balmer decrement, \hb/\hg, with a Calzetti attenuation law \citep{Calzetti2000} adopting $R_V=4.05$.
The observed \hb/\hg\ ratio is $1.921\pm0.174$.
Given that the star formation history inferred from SED fitting suggests a very young stellar population ($\sim1.6$ Myr), we do not expect a significant contribution from underlying stellar Balmer absorption to the measured emission line fluxes.

Assuming Case B recombination with an electron temperature $T_e=22200$ K and an electron density $n_e=500$ cm$^{-3}$, we calculate the intrinsic \hb/\hg\ ratio to be 2.101 using \texttt{PyNeb} \citep{Luridiana2015}.
The adopted temperature is chosen to be consistent with our measurement based on \foiii~$\lambda4363$/\foiii~$\lambda5007$ line ratio (see below), and the electron density is chosen to be consistent with the average of high-redshift galaxies \citep{Hu2024,Li2025}.
The observed ratio is lower than the intrinsic ratio, indicating negligible dust attenuation ($E(B-V)_\mathrm{nebular}<0.03$) in \lae, consistent with the zero dust attenuation inferred from the SED modeling.
Consequently, we do not apply any dust correction in the following analysis.

\vspace{0.5em}
\textit{Electron temperature and density:}
The detection of the \foiii~$\lambda4363$ and \foiii~$\lambda5007$ emission lines allows us to determine the electron temperature in the O$^{2+}$ zone. 
We assume an electron density of $n_e=500$ cm$^{-1}$, consistent with the measurements at similar redshifts \citep[e.g.,][]{Hu2024,Li2025}.
Using the \texttt{getTemDen} function in the \texttt{PyNeb} package, we derive the temperautre $T_e(\mathrm{O^{2+}})=23600^{+1800}_{-1700}$ K. 
We note that \foiii~$\lambda4363$/\foiii~$\lambda5007$ flux ratio is insensitive to the assumed electron density; the \foiii~$\lambda4363$/\foiii~$\lambda5007$ ratio-temperature relation remains nearly unchanged from $n_e=100$ to 10000 cm$^{-3}$.

Due to the non-detection of the \oii\ doublet in \lae, we are unable to directly measure the electron density in the O$^{+}$ zone.

We measure the electron density in the C$^{2+}$ zone using the \ciii~$\lambda\lambda1907,1909$ flux ratio.
Because we do not detect temperature-sensitive C$^{2+}$ emission lines, we adopt the O$^{2+}$ temperature in the C$^{2+}$ zone. 
Although O$^{2+}$ and C$^{2+}$ have different ionization potentials (35 eV and 24 eV, respectively), and the temperatures in the two zones may therefore differ, the \ciii~$\lambda\lambda1907,1909$ ratio shows only a weak dependence on the temperature. This dependence is relatively small compared to the uncertainties in the observed line ratio.
Using \texttt{getTemDen} function in the \texttt{PyNeb} package, we derive the electron density $n_e(\mathrm{C^{2+}})=12900^{+12500}_{-8600}$ cm$^{-3}$.

\vspace{0.5em}
\textit{Oxygen abundance:}
With the electron temperature measured above, we determine O$^{+}$/H$^+$ and O$^{2+}$/H$^+$ by \oii/\hb\ and \foiii/\hb\ ratios, respectively, and we sum O$^{+}$/H$^+ + \textrm{O}^{2+}$/H$^+$ to obtain the total oxygen abundance O/H.
We do not consider the contribution of O$^{3+}$, as it is expected to be minimal even in the highly ionized environments \citep{Berg2019}.
We adopt the $T_e(\mathrm{O^+})$ -- $T_e(\mathrm{O^{2+}})$ relation in \citet{Arellano-Cordova2020} to obtain $T_e(\mathrm{O^+})=14800$ K, and adopt the parametrizations from \citet{Peng2023} to derive the oxygen abundance. 
We derive an upper limit of $12+\log\ (\mathrm{O^+/H})<5.71$ for O$^{+}$ and $12+\log\ (\mathrm{O^{2+}/H})=7.50\pm0.08$ for O$^{2+}$.
The contribution of O$^{+}$ is negligible compared to O$^{2+}$ and is much smaller than the uncertainty of O$^{2+}$/H, therefore, we adopt O$^{2+}$/H as the total oxygen abundance $12+\log\ (\mathrm{O/H})=7.50\pm0.08$, corresponding to $\sim7\%$ of solar metallicity \citep[$12+\log\ (\mathrm{O/H})_\odot=8.69$;][]{Asplund2021}.

\vspace{0.5em}
\textit{Ionization parameter:}
We determine the ionization parameter, $\log U$, from the \foiii/\oii\ line ratio using the calibration of \citet{Berg2019}. 
Adopting the relation for a metallicity of $Z=0.1\,Z_\odot$, we derive a lower limit of $\log U \geq -1.1$ based on the 2$\sigma$ upper limit on the \oii\ flux. 
However, we note that the non-detection of \oii\ could also result from a high electron density ($n_e \gtrsim 3000$ cm$^{-3}$), at which collisional de-excitation significantly suppresses the \oii\ emission.

\vspace{0.5em}
\textit{Carbon-to-Oxygen abundance ratio:}
We determine the carbon-to-oxygen (C/O) abundance ratio using the C$^{2+}$/O$^{2+}$ abundance ratio and an ionization correction factor (ICF): $\mathrm{C/O=C^{2+}/O^{2+}\times ICF}$.
The C$^{2+}$/O$^{2+}$ abundance ratio is calculated using \ciii~$\lambda\lambda1907,1909$/\oiii~$\lambda1666$ ratio and the emissivities of \ciii\ and \oiii\ calculated by \texttt{PyNeb}.
We adopt the electron temperature $T_e=23600^{+1800}_{-1700}$ K and the electron density $n_e=12900^{+12500}_{-8600}$ cm$^{-3}$ for the C$^{2+}$ and O$^{2+}$ zones. 
Accounting for the uncertainties in both the electron temperature and density, we derive $\log\ \mathrm{(C^{2+}/O^{2+})}=-0.92\pm0.09$.

We use the photoionization model-derived ICF from \citet{Berg2019} to correct the contributions of other C ions (mainly C$^+$ and C$^{3+}$). 
\citet{Berg2019} used \texttt{Cloudy} \citep{Ferland2013} and BPASS stellar population model \citep{Eldridge2016,Stanway2016} to calculate the contribution of different C ions to the total abundance as a function of ionization parameter.
Based on $\log U \geq -1.1$, we derive an ICF $\geq1.705$.
The correction is dominated by the contribution of C$^{3+}$, while the contribution of C$^+$ is negligible. 
Applying this ICF to C$^{2+}$/O$^{2+}$, we obtain the C/O abundance ratio $\log\ (\mathrm{C/O})\geq -0.71\pm0.09$.

\section{Discussion} \label{sec:disc}

\subsection{Ionizing Source}
\begin{figure*}
    \centering
    \includegraphics[width=0.48\linewidth]{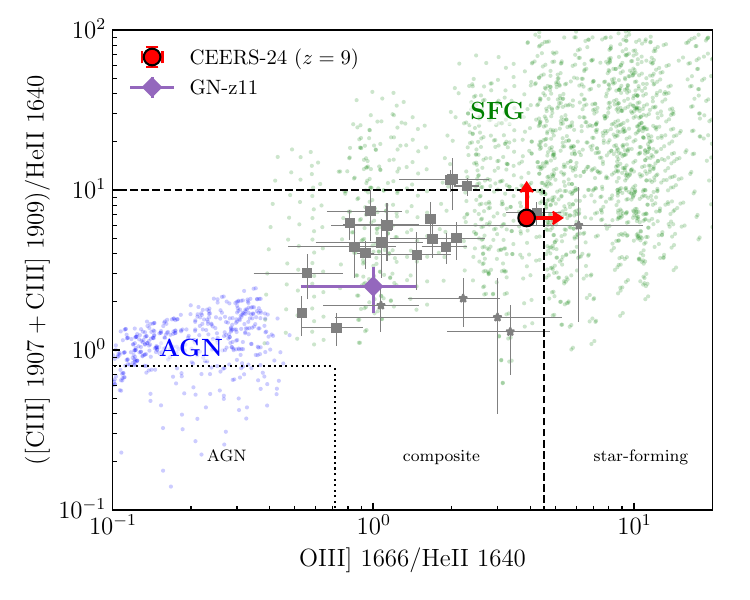}
    \includegraphics[width=0.48\linewidth]{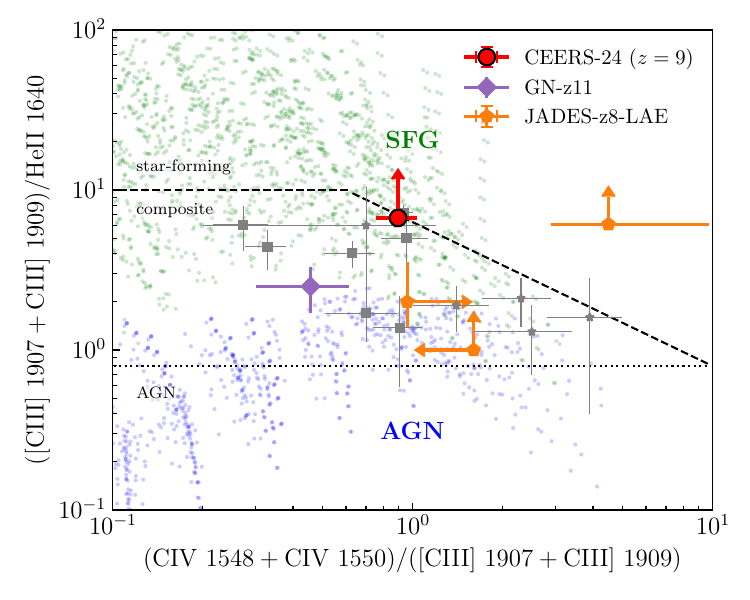}
    \caption{Rest-frame UV emission line diagnostic diagrams. The left panel shows the \ciii~$\lambda\lambda1907,1909$/\heii~$\lambda1640$ -- \oiii~$\lambda1666$/\heii~$\lambda1640$ diagram and the right panel shows the \ciii~$\lambda\lambda1907,1909$/\heii~$\lambda1640$ -- \civ~$\lambda\lambda1548,1550$/\ciii~$\lambda\lambda1907,1909$ diagram.
    The measurements of \lae\ are presented as red dots. 
    In both panels, we collect a series of $z>8$ LAEs and local analog of high-redshift galaxies from the literature, including GN-z11 from \citet{Bunker2023} as purple diamond, three JADES LAEs from \citep{Witstok2025b} as orange pentagons, local analogs from \citet{Jung2025} as gray stars, and local analogs from the CLASSY survey as gray squares \citep{Mingozzi2024}. The blue and green dots in both panels represent the predictions from photoionization models for AGN \citep{Gutkin2016} and star formation \citep{Feltre2016}, respectively.
    The dashed lines and dotted lines are demarcations proposed by \citet{Hirschmann2019} to distinguish between AGN, star-forming galaxies, and AGN--star-forming composite galaxies.}
    \label{fig:diag}
\end{figure*}

The presence of prominent \foiii~$\lambda\lambda4959,5007$ nebular emission lines, together with the non-detection of the \oii\ doublet, implies a hard ionizing radiation field in \lae. 
We estimate an ionization parameter of $\log U > -1.1$ from the \foiii/\oii\ flux ratio (see Section~\ref{sec:ism}), while the SED modeling favors an even higher $\log U \sim 0$.
Additional evidence for a hard ionizing spectrum is provided by the detection of strong high-ionization UV emission lines, such as \civ~$\lambda\lambda1548,1550$. 
Such hard ionizing photons may be produced by intense star formation or by an AGN \citep{ChavezOrtiz2025}. In this section, we investigate the nature of the ionizing sources in \lae.  In the end, we conclude that star-formation dominates the ionization in this galaxy. 

The flux ratios between emission lines with distinct ionization potentials have been frequently used to constrain the hardness of the ionizing source. 
Rest-frame optical diagnostics, such as \foiii~$\lambda5007$/\hb\ to \nii~$\lambda6565$/\ha\ (known as the BPT diagram, \citealp{Baldwin1981}), are among the most widely used and have been extensively tested and calibrated for galaxies at low redshift \citep[e.g.,][]{Kewley2001,Kauffmann2003}.
However, recent studies have demonstrated that these optical diagnostic diagrams lose their discrimination power for identifying ionizing sources of high-redshift galaxies and local metal-poor galaxies \citep[e.g.,][]{Mingozzi2024,Cleri2025,Hirschmann2023}.
Based on the photoionization models calculated by Cloudy \citep{Ferland2013}, \citet{Cleri2025} show that very high-ionization emission lines with ionization potential $>54$ eV (e.g., \heii, \nev) are critical to distinguish between star formation and AGN at high redshift.
Therefore, in this section, we focus on the diagnostic diagrams involving \heii~$\lambda1640$ to constrain the ionizing source in \lae.

Figure \ref{fig:diag} shows the \ciii~$\lambda\lambda1907,1909$/\heii\ $\lambda1640$ versus \oiii~$\lambda1666$/\heii~$\lambda1640$ (hereafter C3He2--O3He2) and \ciii~$\lambda\lambda1907,1909$/\heii~$\lambda1640$ versus \civ~$\lambda\lambda1548,1550$/\ciii~$\lambda\lambda1907,1909$ (hereafter C3He2--C4C3) diagrams. 
We present our measurements for \lae\ along with those from other $z>8$ LAEs \citep{Bunker2023,Witstok2025b} and local metal-poor dwarf galaxies \citep{Mingozzi2024,Jung2025} for comparison.
Due to \heii\ not detected in the spectrum, we adopt its 2$\sigma$ upper limit. 
\lae\ shows very large \ciii/\heii\ and \oiii/\heii\ line ratios compared to the local dwarf galaxies and other $z>8$ LAEs, indicating that \lae\ produces less hard ionizing photons with $>54$ eV than those galaxies. 

We then compare our measurements with the theoretical predictions from the photonionization models for AGN \citep{Feltre2016} and star formation \citep{Gutkin2016}.
We consider the models with dust-to-metal mass ratios in the range 0.1 -- 0.5, hydrogen densities in the range 10 -- $10^4$ \cmcu, ionization parameters in the range $-4$ to $-1$, and metallicities in the range 0.001 -- 0.004 (corresponding to 0.066 -- 0.26 $Z_\odot$ based on $Z_\odot=0.01524$ assumed in \citealp{Feltre2016}). 
We find that \lae\ is consistent with the predictions of star formation models but significantly deviates from those of AGN models, as shown in Figure \ref{fig:diag}. 
In addition, we compare \lae\ with the demarcations proposed by \citet[][dashed and dotted line in Figure \ref{fig:diag}]{Hirschmann2019}, which coupled these two models with cosmological zoom-in simulations.
\lae\ locates on the borderlines between the SF region and SF-AGN composite region in both C3He2--O3He2 and C3He2--C4C3 diagrams, ruling out AGN as the dominant source of ionization. 

Consequently, we conclude that star formation is the dominant source of ionizing photons in \lae.

\subsection{\lya\ Escape and Ionized Bubble Surrounding \lae} \label{sec:lya}

\begin{figure*}
    \centering
    \includegraphics[width=\linewidth]{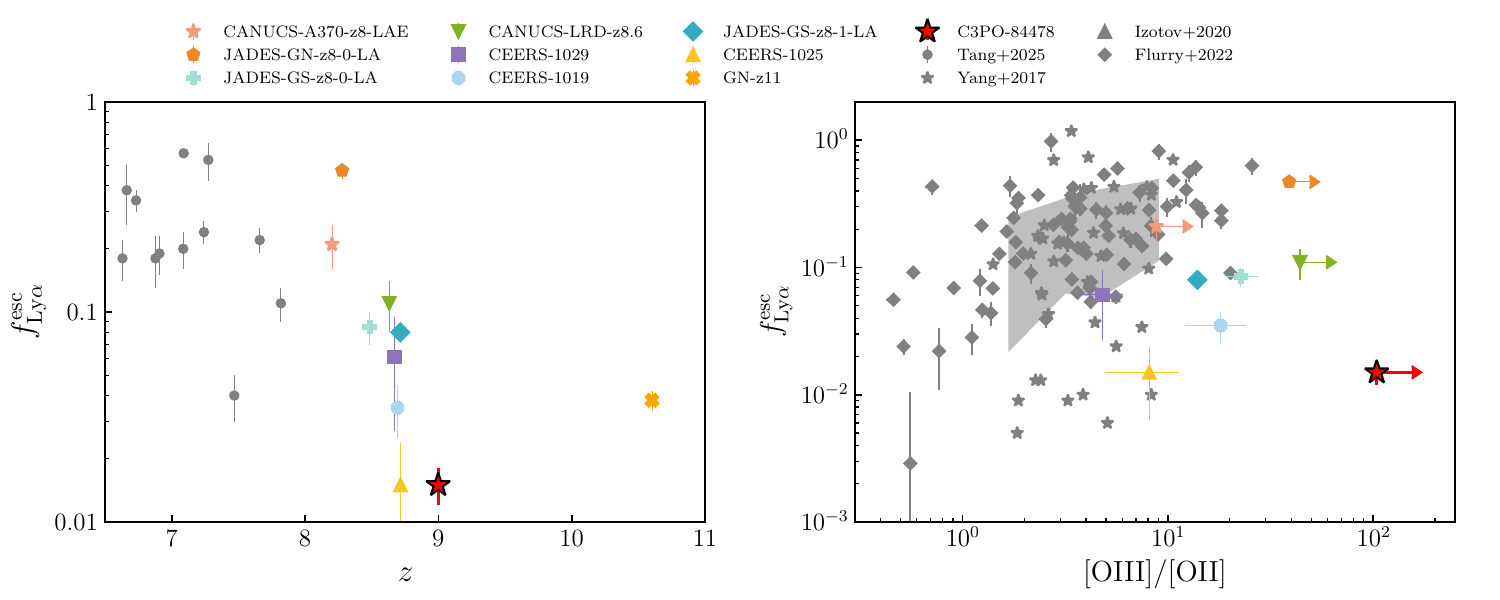}
    \caption{\lya\ escape fraction as a function of redshift (left) and \foiii/\oii\ ratio (right). The measurements of \lae\ are marked as red stars. 
    For comparison, we collect LAEs at $8<z<11$ from the literature, including GN-z11 \citep{Bunker2023}, CEERS-1019 \citep{Tang2023}, CEERS-1029 \citep{Larson2023,Tang2023}, CEERS-1025 \citep{Whitler2025}, three JADES LAEs \citep{Witstok2025b}, and two CANUCS LAEs \citep{Willott2025,Morishita2025}. We include the LAEs at $6.5<z<8$ in the CEERS, JADES, and A2744 fields, adopting the measurements from \citet{Tang2024}. We also plot the green pea galaxies from \citet{Yang2017} as gray stars and local LyC leakers from \citet{Izotov2020} and \citet{Flury2022} as gray triangles and gray diamonds.
    The gray shaded region in the right panel represents the $f^\mathrm{Ly\alpha}_\mathrm{esc}$--\foiii/\oii\ relation derived from local LyC leakers by \citet{Hayes2023}.}
    \label{fig:fesc_z}
\end{figure*}

The shutters for \lae\ were positioned near the lower center of NIRSpec Quadrant 2, causing the wavelength interval 11167.05 -- 12149.37 \AA\ to fall within the detector gap.
Fortunately, we detect an emission line at 12162.5~\AA, lying just four spectral pixels beyond the edge of the gap. 
This line is consistent with \lya\ at $z = 8.99781$ with a velocity offset of 210~\kms\ from systemic. 
Given that intrinsic \lya\ line profiles are often highly asymmetric and that the G140M spectral resolution cannot fully resolve this structure, we conservatively adopt half a spectral pixel as the velocity uncertainty, corresponding to $\sim80$~\kms.

For clarity, we define the intrinsic \lya\ escape fraction $f^\mathrm{int,esc}_\mathrm{Ly\alpha}$ as the fraction of \lya\ photons escaped from the galaxy. This describes the effects of resonant scattering of neutral hydrogen and dust absorption of ISM and CGM. 
To be consistent with other high-redshift studies, we define the \lya\ escape fraction $f^\mathrm{esc}_\mathrm{Ly\alpha}$ as the fraction of \lya\ photons observed by us. This incorporates both the intrinsic \lya\ escape fraction and the IGM transmission: $f^\mathrm{esc}_\mathrm{Ly\alpha}=f^\mathrm{int,esc}_\mathrm{Ly\alpha}\times T_\mathrm{IGM}$.

We calculate the \lya\ escape fraction $f^\mathrm{esc}_\mathrm{Ly\alpha}$ by the observed \lya\ flux to intrinsic \lya\ flux ratio.
Assuming Case B recombination and no dust, we infer the intrinsic \lya\ flux from the observed \hb\ flux using $F^\mathrm{int}_\mathrm{Ly\alpha}=24.77\times F_\mathrm{H\beta}$ and derive a \lya\ escape fraction of $f^\mathrm{esc}_\mathrm{Ly\alpha}=1.5\pm0.3\%$.
Figure \ref{fig:fesc_z} presents the \lya\ escape fraction of \lae, along with those of reionization-era galaxies as a function of redshift. 
We collect the \lya\ escape fraction measurements at $z>8$ from literature, including GN-z11 \citep{Bunker2023}, CEERS-1019 \citep{Tang2023}, CEERS-1029 \citep{Larson2023,Tang2023}, CEERS-1025 \citep{Whitler2025}, three JADES LAEs \citep{Witstok2025b}, and two CANUCS LAEs \citep{Willott2025,Morishita2025}.
We do not include JADES-GS-z13-1-LA, because its \lya\ escape fraction is not robustly measured yet.
We further include the LAEs at $6.5<z<8$ observed by the NIRSpec Grating spectroscopy in the CEERS \citep{Finkelstein2025}, JADES \citep{Eisenstein2023,Eisenstein2023b}, and A2744 \citep{Bezanson2024} surveys, adopting the measurements from \citet{Tang2024}.

\lae\ exhibits a \lya\ escape fraction significantly lower than typical galaxies at $6.5 < z < 8.3$, but comparable to those at $z > 8.3$, which generally have $f_{\mathrm{Ly\alpha}}^{\mathrm{esc}} \sim 1\%$ -- 10\%. 
The rapid decline in \lya\ escape fraction with increasing redshift reflects the rising neutral hydrogen fraction of the IGM. 
The \lya\ escape fraction of \lae\ is among the lowest of the galaxies at $z>8$, at odds with its exceptionally high \foiii/\oii\ ratio and high surface star formation rate, which are thought to probe high ionizing photon escape and \lya\ escape \citep[e.g.,][]{Yang2017,Flury2022}.
This could indicate that \lae\ resides in a relatively small ionized bubble where its \lya\ photons are largely attenuated by the IGM.

We estimate the required size of the ionized bubble surrounding \lae\ to explain the observed \lya\ escape fraction using the prescription proposed in \citet{Mason2020} (see also \citealp{Witstok2024} and \citealp{Martin2025}).
We assume that the galaxy resides in the center of an ionized bubble and the IGM is completely neutral outside the bubble. 
The neutral IGM outside the bubble and the residual neutral hydrogen within the bubble together contribute to the \lya\ attenuation.
Considering the recombination within the ionized bubble, the residual neutral hydrogen fraction is assumed to be $\chi_\mathrm{HI}\propto r^2$, where $\chi_\mathrm{HI}(0.1~\mathrm{pMpc})=10^{-8}$.

We first consider a simplified model where the \lya\ photons are free to escape from the galaxy, such that all \lya\ attenuation is attributed to the IGM. 
We assume that the emergent \lya\ line from the galaxy peaks at the observed \lya\ line center, corresponding to a velocity offset of $\Delta v_{\mathrm{Ly\alpha}} = 210$ \kms.
To allow an IGM transmission of $1.5\pm0.3\%$ at $\Delta v=210$ \kms, we derive a bubble radius of 0.073 pMpc. 
This estimation can be regarded as the lower limit of the required bubble radius because \lya\ line is usually asymmetric and extends towards larger velocities.
Additionally, as argued in \citet{Witstok2024}, other effects, such as slitloss, dust absorption, and smaller \lya\ velocity offset, can result in an overestimation of IGM transmission. 

We then consider a model taking ISM attenuation into account, i.e., intrinsic \lya\ escape fraction $f^\mathrm{int,esc}_\mathrm{Ly\alpha}<1$
Studies of low-redshift galaxies have revealed that intrinsic \lya\ escape fraction correlates with several galaxy properties, including the galaxy mass, metallicity, and \foiii/\oii\ line ratio \citep[e.g.,][]{Yang2017,Hayes2023}.
Specifically, \citet{Hayes2023} have presented a strong correlation between \lya\ escape fraction and \foiii/\oii\ ratio based on the \hst\ Cosmic Origins Spectrograph observations of 87 low-redshift galaxies.
For the galaxy sample with $\log$ \foiii/\oii\ $\sim0.92$ -- 1.82, \citet{Hayes2023} derived an escape fraction of $\sim25\%$ for the red peak.
Adopting this escape fraction as the intrinsic \lya\ escape fraction of \lae, we derive a $T_\mathrm{IGM}=6.0\pm1.2\%$.
Using this IGM transmission, we derive a bubble radius of $0.196\pm0.025$ pMpc.

Therefore, based on the different methods, the bubble radius for \lae\ is approximately between 0.07 to 0.20 pMpc.

Can \lae\ produce such an ionized bubble?
We use the Str\"omgren sphere model \citep{Cen2000} to estimate the size of the bubble that can be produced solely by \lae.
The growth of the ionized bubble is described by:
\begin{equation} \label{eq:1}
    \frac{dR^3_i}{dt} = 3 H(z)R^3_i + \frac{3\dot N_\mathrm{int} {f^\mathrm{esc}_\mathrm{LyC}}}{4\pi n_\mathrm{H}} - C n_\mathrm{H} \alpha_\mathrm{B} R^3_i,
\end{equation}
where the three terms correspond to the Hubble expansion, the ionization by the central galaxy, and the recombination of the ionized bubble, respectively. 
The ionization term is determined by the intrinsic production rate of ionizing photons $\dot N_\mathrm{int}$ and ionizing photon escape fraction $f^\mathrm{esc}_\mathrm{LyC}$.
Under Case B recombination \citep{Storey1995,Osterbrock2006}, we can infer the ionizing photon production rate from \hb\ luminosity via:
\begin{equation}
    \dot N_\mathrm{int} = 2.10 \times 10^{12} \times \frac{L_\mathrm{H\beta}}{1-f^\mathrm{esc}_\mathrm{LyC}}.
\end{equation}

Directly measuring $f^\mathrm{esc}_\mathrm{LyC}$ is infeasible for galaxies in the epoch of reionization. 
Instead, we adopt the value $f^\mathrm{esc}_\mathrm{LyC}=25^{+3}_{-4}\%$ inferred from the SED fitting (see Section \ref{sec:sed}). 
For the duration of ionizing photon production, we adopt the age of the constant star formation history.
Assuming that \lae\ produces ionizing photons continuously over this period, we integrate Equation \ref{eq:1} to estimate a bubble radius of $0.095^{+0.005}_{-0.007}$ pMpc.

In addition, numerous empirical relations between ionizing photon escape fraction and galaxy properties have been explored for low-redshift galaxies \citep[e.g.,][]{Flury2022,Chisholm2022}. 
For example, a high \foiii/\oii\ ratio may indicate density-bounded H {\sc ii} regions through which ionizing photons can escape \citep{Nakajima2014}, while a blue UV slope could reflect less dust attenuation and small contribution of nebular emission \citep{Zackrisson2013,Chisholm2022}.
\citet{Jaskot2024a,Jaskot2024b} used the survival analysis technique to generate multivariate models for predicting $f^\mathrm{esc}_\mathrm{LyC}$ based on the Low-redshift Lyman Continuum Survey, and found that UV slope ($\beta_\mathrm{UV}$), star formation surface density, and \foiii/\oii ratio are among the top-ranked variates.
Applying the ``JWST model'' of \citet{Jaskot2024a}, which incorporates these three parameters, to \lae, we infer a $f^\mathrm{esc}_\mathrm{LyC}=0.32^{+0.32}_{-0.21}$, consistent with that inferred from the SED fitting.
This implies the bubble radius produced by \lae\ approximates $0.107^{+0.060}_{-0.048}$ pMpc.

In summary, our estimates of the bubble radius produced by \lae\ are comparable to the minimum ionized bubble radius required for its \lya\ transmission, suggesting that \lae\ alone is capable of producing such a bubble.

\subsection{Environment of \lae}

\begin{figure*}[htbp]
    \centering
    \includegraphics[width=\linewidth]{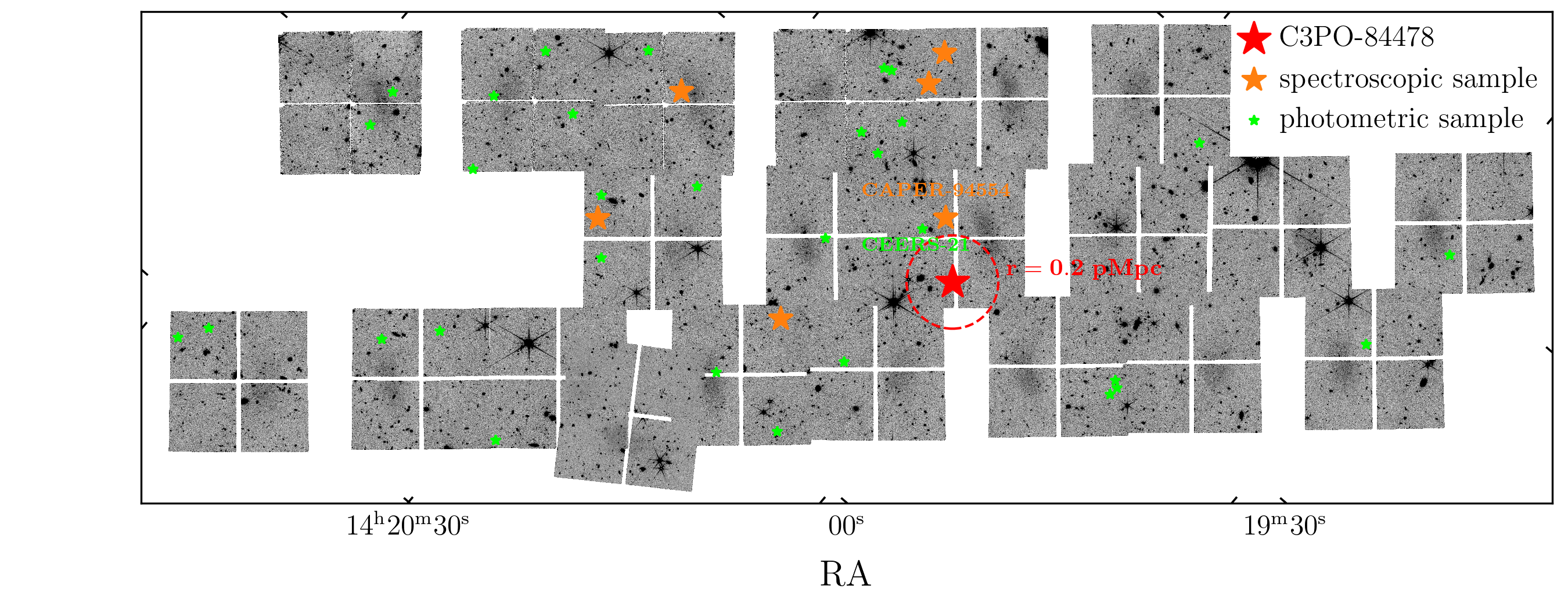}
    \caption{Location of \lae\ (red star) within the CEERS NIRCam footprint. We collect spectroscopically confirmed galaxies at $8.9<z<9.1$ from the CEERS, RUBIES, CAPERS, and GO-4287 programs, and present them as the orange stars. We also present photometric samples at $8.7<z<9.3$ from \citet{Finkelstein2024} as green stars. The red dashed circle with $r=0.2$ pMpc represents the size of the ionized bubble estimated in Section \ref{sec:lya} and no companion galaxy is identified within this circle. However, we identify two galaxies CEERS-21 and CAPERS-EGS-94554 at a projected distance of $<$ 0.3~pMpc.}
    \label{fig:env}
\end{figure*}

In this section, we explore the environment of \lae, as galaxies residing in overdense regions can collectively produce sufficiently large ionized bubbles compared to isolated field galaxies, thereby alleviating the attenuation of \lya\ photons by the neutral IGM \citep{Qin2022,Hutter2023,Chen2026}. 

Several studies have reported a galaxy overdensity at $z\sim8.7$ in the EGS field \citep{Larson2022, Whitler2025}.
This overdensity includes three confirmed LAEs, suggesting the presence of large ionized bubbles \citep{Whitler2025}. 
However, \lae\ is unlikely to be associated with this overdensity, because the line-of-sight distance between \lae\ and the overdensity is $\gtrsim 7$ pMpc, much larger than the bubble radius at $z\sim9$ suggested by the theoretical predictions \citep[$<2$ pMpc;][]{Lu2024}.

Figure \ref{fig:env} shows the location of \lae\ within the CEERS NIRCam footprint \citep{Bagley2023}, along with the spectroscopic sample and photometric sample at similar redshifts. 
We collect the spectroscopically-confirmed galaxies at $8.9<z_\mathrm{spec}<9.1$ from the CEERS \citep{Finkelstein2025}, RUBIES \citep{deGraaff2025}, CAPERS \citep{Dickinson2024}, and \jwst\ GO-4287 \citep{Mason2023} programs. 
We also include photometrically-selected galaxies at $8.7<z_\mathrm{phot}<9.3$ from \citet{Finkelstein2024}. 
The adopted redshift interval of $\Delta z=0.6$ accounts for the typical uncertainty in photometric redshift measurement. 
We exclude three photometrically-selected galaxies that are confirmed to be low-redshift interlopers by spectroscopy.

As shown in Figure~\ref{fig:env}, we find no evidence for a large-scale galaxy overdensity in the vicinity of \lae.
We also do not identify any companion galaxies within a projected distance of 0.2~pMpc, corresponding to the inferred ionized bubble radius of \lae.
However, we identify two galaxies within a projected distance of 0.3~pMpc: CEERS-21 and CAPERS-EGS-94554.

CEERS-21 has a photometric redshift of $z_\mathrm{phot}=9.01\pm0.30$. It was observed with \jwst\ NIRSpec medium gratings as part of the CEERS survey; however, no robust emission line was detected, and the spectroscopic redshift remains unconstrained. 
If CEERS-21 is physically associated with \lae, we can combine its NIRCam photometry and NIRSpec spectroscopy to estimate the equivalent width of \foiii~$\lambda5007$. We obtain a $2\sigma$ upper limit of $\mathrm{EW_0}<160$ \AA. 
This relatively small \foiii\ equivalent width suggests a low level of star formation activity, indicating that CEERS-21 cannot contribute significantly to the ionization of the IGM surrounding \lae.

CAPERS-EGS-94554 lies at a projected separation of $0.28$ pMpc from \lae.
It was observed with \jwst\ NIRSpec prism as part of the CAPERS survey, and its redshift has been confirmed to be $z_\mathrm{spec}=9.012$.
The small redshift difference between \lae\ and CAPERS-EGS-94554 (corresponding to 425 \kms) could arise from the peculiar velocities of the two galaxies; thus, CAPERS-EGS-94554 could reside in the same ionized bubble as \lae.
CAPERS-EGS-94554 has a very large \foiii+\hb\ equivalent width of $\mathrm{EW_0}=1250\pm110$ \AA\ and a large \foiii/\oii\ ratio of $>8.5$, suggesting that CAPERS-EGS-94554 may be a LyC leaker.
However, the \hb\ luminosity of CAPERS-EGS-94554 is 25\% of \lae, indicating that even if CAPERS-EGS-94554 has a similar ionizing photon escape fraction, its contribution to the ionizing budget of the ionized bubble is much smaller than \lae.

Overall, we find no evidence that nearby galaxies dominate the local ionizing photon budget, supporting the conclusion that \lae\ is the primary source responsible for producing the ionized bubble required for the transmission of its observed \lya\ emission.

\section{Summary} \label{sec:summ}

We report new \jwst\ NIRSpec G140M and G395M observations from the C3PO program for a previously known galaxy at $z=8.99781$, \lae.
We detect a emission line at 12162.5 \AA, corresponding to \lya\ line of \lae\ with a velocity offset of $\Delta v_{\mathrm{Ly\alpha}}=210\pm80$ \kms.
This makes \lae\ the third LAE confirmed at $z\gtrsim9$.
Our main results are as follows:
\begin{itemize}
    \item \lae\ is a low-metallicity, low-mass galaxy. 
    By jointly modeling photometry and spectrum of \lae, we infer a stellar mass of $M_\star = 2.2\pm0.1 \times 10^7\ M_\odot$.
    Based on the temperature measured from the aurora line \foiii~$\lambda4363$, we derive the oxygen abundance of $12+\log(\mathrm{O/H}) = 7.50\pm0.08$ and carbon-to-oxygen abundance ratio of $\log\ \mathrm{(C/O)} = -0.71\pm0.09$. 
    \lae\ is very compact and is dominated by a recent burst of star formation with a high star formation rate surface density of $\geq 142^{+25}_{-23}\ M_\odot$ yr$^{-1}$ kpc$^{-2}$ and an age of $1.6\pm0.2$ Myr.
    \item We explore the ionizing source of \lae\ using the UV line diagnostic diagrams, C3He2--O3He2 and C3He2--C4C3. These diagnostic diagrams disfavor \lae\ to be powered by AGN, and indicate that star formation is the primary ionizing source.
    \item \lae\ shows a small \lya\ escape fraction of $1.5\pm0.3\%$. To explain this observed \lya\ escape fraction, \lae\ should reside in an ionized bubble with a radius of $R\sim 0.07$ -- 0.20 pMpc. The high \foiii/\oii\ ratio, high star formation rate surface density, and blue UV slope of \lae\ indicate that \lae\ is a strong ionizing photon leaker. Adopting the ionizing photon escape fractions inferred from both the SED fitting and empirical calibrations, we find that \lae\ alone is capable of producing the required ionized bubble.
    \item We find no evidence for a large-scale galaxy overdensity around \lae\ and identify no companion galaxies within the inferred bubble radius ($<0.2$ cMpc). Although a spectroscopically confirmed galaxy, CAPERS-EGS-94554 ($z=9.012$), lies at a projected separation of only $\sim0.28$~pMpc, its contribution to the ionizing photon budget is likely much smaller than that of \lae.
\end{itemize}

Our results imply that we could be witnessing the early formation of an ionized bubble driven by a low-mass vigorous star-forming galaxy during the onset of cosmic reionization.  
Similar systems may be common at these epochs; however, their low \lya\ escape fractions likely render them difficult to detect. 
Our observation demonstrates that future deep G140M or G140H spectroscopic observations will be essential to uncover this population and to identify ionized bubbles at their earliest stages.

\begin{acknowledgments}
This work is based on observations made with the NASA/ESA/CSA James Webb Space Telescope. The data were obtained from the Mikulski Archive for Space Telescopes at the Space Telescope Science Institute, which is operated by the Association of Universities for Research in Astronomy, Inc., under NASA contract NAS 5-03127 for JWST. These observations are associated with program \#5943.  Support for US investigators in program \#5943 was provided by NASA through a grant from the Space Telescope Science Institute, which is operated by the Association of Universities for Research in Astronomy, Inc., under NASA contract NAS 5-03127.
\end{acknowledgments}


\facilities{\textit{JWST}}
\software{astropy, BAGPIPES, PyNeb}

\appendix
\section{Best-fit SED of \lae}

\renewcommand{\thefigure}{A1}
\begin{figure*}
    \centering
    \includegraphics[width=0.9\linewidth]{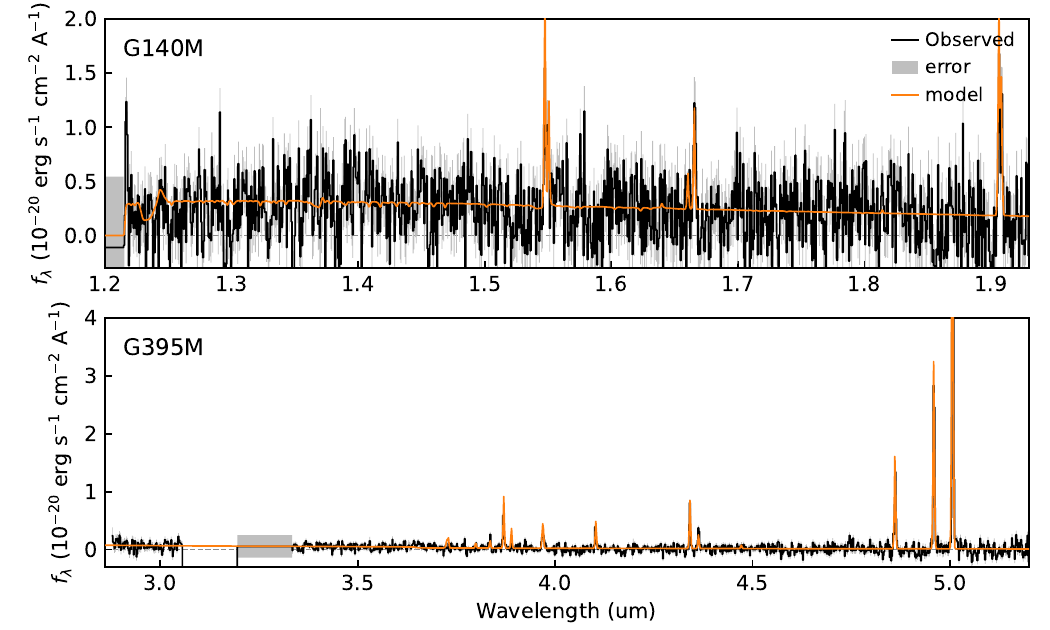}
    \caption{The best-fit SED of \lae\ using \texttt{BAGPIPES} with the ionizing photon escape model. The black curves represent the 1D spectra and the gray shades represent the error spectra. The solid orange curves represent the best-fit model.}
    \label{fig:bestsed}
\end{figure*}

\bibliography{main}{}
\bibliographystyle{aasjournalv7}



\end{document}